\documentclass[12pt,preprint]{aastex}
\usepackage{latexsym}
\usepackage{amssymb}
\usepackage{natbib}

\newcommand{\marc}{mag~arcsec$^{-2}$}
\newcommand{\mue}{$\mu_{\rm e}$~}
\newcommand{\re}{$R_{\rm e}$~}

\begin{document}

\bibliographystyle{apj}

\title{1.65 $\mu$m ($H$-band) surface photometry of galaxies. X: Structural and dynamical properties of elliptical galaxies.}

\author{Stefano Zibetti\altaffilmark{1}}

\affil{Max-Planck-Institut f\"ur Astrophysik Karl-Schwarzschild-Str. 1, D-85741 Garching bei M\"unchen, Germany}

\altaffiltext{1}{Universit\`a degli Studi di Milano - Bicocca, P.zza 
delle Scienze 3, I-20126 Milano, Italy}

\email{zibetti@mpa-garching.mpg.de}

\and 

\author{Giuseppe Gavazzi}

\affil{Universit\`a degli Studi di Milano - Bicocca, P.zza 
delle Scienze 3, I-20126 Milano, Italy}

\and 

\author{Marco Scodeggio}

\affil{Istituto di Fisica Cosmica ``G. Occhialini'', CNR, via Bassini 15, 
I-20133, Milano, Italy}

\and

\author{Paolo Franzetti\altaffilmark{1}}

\affil{Istituto di Fisica Cosmica ``G. Occhialini'', CNR, via Bassini 15, 
I-20133, Milano, Italy}

\and

\author{Alessandro Boselli}

\affil{Laboratoire d'Astrophysique de Marseille, BP8, Traverse du Siphon, F-1337
6 Marseille 
Cedex 12, France}

\begin{abstract}We analyse the structural and dynamical
properties of a sample of 324 nearby elliptical and dwarf elliptical galaxies observed
during an extensive NIR survey in H-band (1.65\micron). The Fundamental Plane (FP) is
determined and a significant tilt is assessed. The origins of such a tilt are investigated
by means of a spherically symmetric, isotropic pressure supported dynamical model
relying on the observed surface brightness profiles. The systematic variation
of the shape coefficient converting the measured central velocity dispersion $\sigma_0$ into
the virial rms velocity $\sigma_{rms}$ is found to be the main cause of the tilt, due
to aperture effects. Moreover
the ratio between the dynamical mass $M_{dyn}$ and the total H-band luminosity $L_H$
turns out to be roughly constant along the luminosity sequence of ellipticals: H-band
luminosity is therefore a reliable and cheap estimator of the dynamical mass of the Es.
\end{abstract}
\keywords{galaxies: fundamental parameters --- galaxies: elliptical and lenticular, cD --- galaxies: photometry --- infrared: galaxies}


\shorttitle{H-band Fundamental Plane}

\shortauthors{S. Zibetti et al.}

\section{Introduction}
The assumption that normal galaxies are in dynamical equilibrium implies
that, for a given type of dynamics (i.e. rotation or pressure supported),
the dynamical status of the system is strongly related to its mass distribution.
Using the further assumption that the mass inside the ``observable'' radius
of a galaxy is traced by the (stellar) light, the dynamical parameters 
should be, at least in first approximation, determined by the structural parameters 
describing the light distribution.
In the case of elliptical galaxies, these systems have been proved to be mainly
pressure supported and the fundamental dynamical parameter
is the central velocity dispersion $\sigma_0$. The simplest way of describing the
light distribution of a galaxy is to measure the half light (or effective) radius \re and
the average surface brightness $I_e$ inside \re. If the elliptical galaxies
formed a homologous family, both from the structural and the dynamical point of view,
and the $M/L$ ratio were constant, the virial theorem
would imply a linear relation between the logarithm of the three parameters
given by:
\begin{equation}
\mathrm{Log}~R_e=2~\mathrm{Log}~ \sigma_0 - \mathrm{Log}~ I_e + k
\end{equation}
A linear relation has been actually found \citep[the Fundamental Plane, FP,][]{ddFP,dresslerFP},
but the coefficients differ significantly from those predicted by the virial 
theorem (this is known as the ``tilt'' of the FP). This implies the 
non-constancy of $M/L$ and/or the breaking of the homology. 
However the existence of the FP as a tight relation requires that the
variations of $M/L$ or the breaking of the homology happen in a very
systematic way.\\ 
Most of the early studies devoted to this problem assumed homology and concluded that a
systematic increase of $M/L$ with luminosity is needed both in optical \citep[see e.g.][]{k3} and
in NIR pass-bands \citep[see e.g.][]{pahre98b}.
However, many studies, in which the surface brightness profiles are analysed both in
optical \citep[see e.g.][]{djorg_etal85,1985nagp.meet..257D,
1986ApJS...60..603S,1987ApJS...64..643S,1987MNRAS.226..747J,1988ApJS...68..173D,
1988AJ.....96..487C} and in NIR pass-bands 
\citep[see e.g.][]{dEvirgo,c31marco}, have found systematic variations of 
the profile shape and concentration index (i.e. the structural parameter quantifying
how much the light distribution is centrally peaked\footnote{Many definitions of
concentration index were proposed in literature; the definition
hereafter adopted is $c_{\mathrm 31}=r_{75}/r_{25}$, the ratio between the radii enclosing
75\% and 25\% of the total luminosity of the galaxy \citep[see][]{dEvirgo,
c31marco}})
among ellipticals, implying a breaking of the homology.
In pioneering work \cite{prugniel97} and \cite{graham97} produced evidence
in favour of a strong influence of structural non-homology on the tilt of the FP.
\cite{busarello97} concluded that most of the tilt could be accounted for by dynamical
non-homology, although spatial non-homology and stellar population effects can
give significant contributions as well.
More recently, \cite{Bertin02} showed how the departure from spatial homology contributes 
to the tilt of the FP in the B band for a small sample of nearby E galaxies imaged
with high S/N.\\
The claim by \cite{FPmarco}, that the amount of tilt is significantly lower in the 
NIR bands than in the optical ones, implies, however, a significant role of different
stellar populations (age, metallicity) in determining the M/L and the structural parameters.
NIR pass-bands are the most suitable for studying the structural parameters 
because they trace much better than visible pass-bands the bulk of the luminous mass 
of the galaxies which sits in old stellar populations. Moreover they are less sensitive 
to dust obscuration and line blanketing, and in turn this reduces the age/metallicity 
effects on the $M/L$ ratio.
We have performed a 
systematic and extensive investigation of the structural properties in the H (K')
band for a sample of nearby galaxies, covering all the morphological types and
extending from high to low luminosities, i.e. to the dwarf regime 
\citep[see][~and references therein]{paperV,dEvirgo}. Based on these
surface brightness profiles, \cite{c31marco} demonstrate a systematic 
relationship between the concentration index $c_{31}$
and the total H-band luminosity which is almost completely independent from the
eye-ball morphological classification: from this relationship the structural non-homology of the
elliptical family can be inferred.
The structural and dynamical properties of the whole sample of surveyed galaxies
are analysed by \cite{pieriniKspace} adopting the $\kappa$-space formalism \citep{k3}.\\
In this paper we present a study of the relationships between the structural 
H-band (1.65\micron) properties of 324 elliptical and dwarf elliptical galaxies.
The relationships with dynamical parameters are analysed for a subsample of 135 
galaxies ranging from the highest luminosities to the dwarf regime and extending
the original sample of 73 galaxies studied by \cite{FPmarco}.
The sample
is described in Sec. \ref{sample}. In Sec. \ref{kormendy_sect} we study the
``Kormendy relation'' between the effective surface brightness
\mue and the effective radius \re. In Sec. \ref{FP_sect} a derivation of the
H-band FP is presented using different fitting methods, with an analysis of
the contribution to the FP tilt due to $M/L$ and homology breaking is
performed in Sec. \ref{tilt_sect}, using a simple dynamical model relying on
the measured surface brightness profiles. 
The method, independently developed by \cite{Zibetti_tesi}, is very
similar to that of \cite{Bertin02}. It should be stressed, however, that we analysed a 
10 times larger sample using NIR photometric data, as opposed to the B band data
analysed by \cite{Bertin02}.
A brief discussion
and the conclusions of this work are given in Sec. \ref{discussion} and
\ref{conclusions}.

\section{The sample}\label{sample}

In this paper we analyse the photometric H-band (1.65\micron)
structural parameters of a sample of 324 ``bona fide'' elliptical and
dwarf elliptical (dE) galaxies selected among members of 5
nearby, rich clusters: namely the Virgo, Coma, A1367, A262 and Cancer
clusters, in addition to a significant population of galaxies in the
``Great Wall'', the bridge between Coma and A1367. Observations and photometric 
analysis of this sample are reported by 
\cite{paperIII,paperV,dEvirgo} (and references therein). The total H-band
luminosity $L_H$, the effective radius \re (i.e. the radius
enclosing half of the total luminosity), the effective surface
brightness \mue, defined as the average surface brightness within \re,
and the concentration index $c_{31}$ are derived from the
azimuthally averaged surface photometry \citep[see][]{paperV}.
Absolute luminosities and physical radii are calculated attributing to each galaxy
the mean cluster (``cloud'') distance, with the exception of the galaxies in the
``Great Wall'', for which the redshift distance ($\rm H_{\rm 0}=75~km~sec^{-1}~Mpc^{-1}$)
is assumed.\\
For a subsample of 135 galaxies dynamical analysis could be performed relying
on central velocity dispersion measures ($\sigma_0$) obtained from
the literature \citep[see][~and references therein]{FPmarco}.\\
The entire sample comprises:
274 ellipticals (129 with $\sigma_0$ measurements) and
50 dwarf ellipticals/spheroidals (6 with $\sigma_0$ measurements).\\ 

\clearpage
\begin{table}[ht]
\caption{Completeness for the Coma Super-cluster and the Virgo cluster}
\label{completeness}
\[
\begin{array}{l|c|c|c|cc|cc}
\hline
Region&m_{\rm pg}& \mathrm{Log}~L_{\rm H\odot}& Tot&\multicolumn{2}{c|}{Photo}&
\multicolumn{2}{c}{Dynamic}\\
&&&\#&\#&\%&\#&\%\\
\hline
Coma&<15.7&>10.5&186&180&(97)&86&(46)\\
Virgo&<14&>9.6&35&31&(89)&25&(71)\\
&14\div15&9.6\div9.1&72&36&(50)&7&(10)\\
&15\div16&9.1\div8.7&99&12&(12)&2&(2)\\
&16\div18&8.7\div7.8&304&6&(2)&0&(0)
\end{array}
\]
\end{table}
\clearpage

Details on the sample completeness in the Coma Super-cluster
region ($\rm 18^o \le \delta \le 32^o$; $\rm 11^h30^m \le \alpha \le 13^h30^m$) 
and the Virgo cluster are given in Tab.\ref{completeness}: for each bin of 
photographic magnitude $m_{\rm pg}$ ($\mathrm{Log}~L_{\rm H\odot}$) the total number
of galaxies in the CGCG \citep[][  $m_{\rm pg}<15.7$]{CGCG} -- for the Coma region -- and 
in the VCC catalogue \citep{VCC} -- for the Virgo cluster -- are given, along with
the total number (percentage) of galaxies in the photometric sample
and in the dynamical sample. In translating $m_{\rm pg}$ into 
$\mathrm{Log}~L_{\rm H\odot}$
an average colour-luminosity relation of $B-H=0.3 \mathrm{Log}~L_{\rm H\odot}+0.5$ is assumed 
\citep[see][]{c31marco}. A mean distance modulus of 34.9 is assumed for the Coma
region, while a distance of 17 Mpc is attributed to the Virgo A cluster, according
to \cite{G99}.\\
As far as the photometric parameters are concerned, we cover a 
representative sample comprising the transition between the giant- and the dwarf-regime.
Dynamical measurements are available only for giants, with few exceptions
providing some hints on the 
behaviour of the scaling relations at low luminosity.
\section{The Kormendy relation}\label{kormendy_sect}

\clearpage
\begin{figure}
\plotone{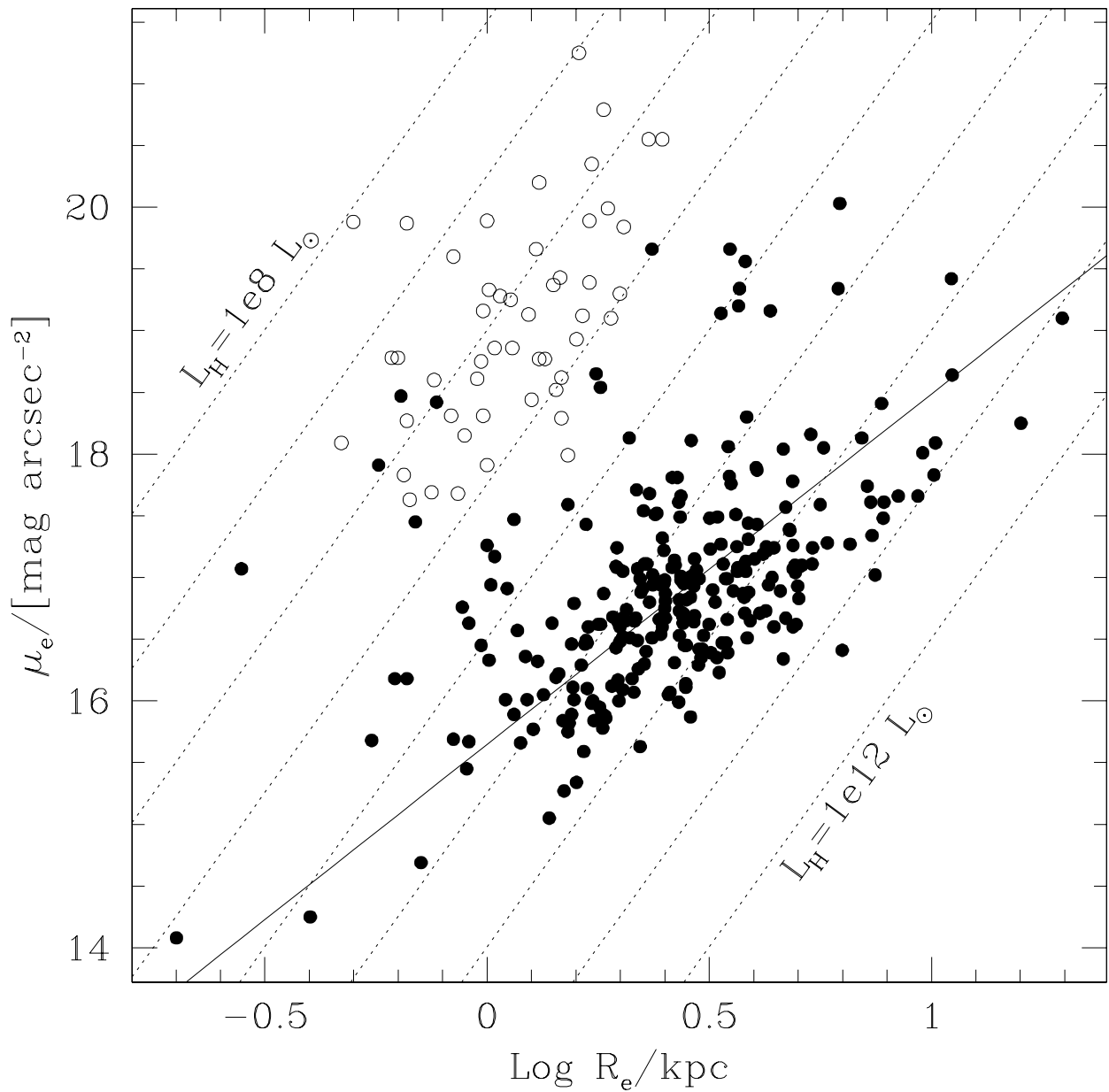}
\caption{The Kormendy relation. Filled symbols are giant ellipticals, empty symbols
dwarfs. Iso-luminosity lines are dotted, assuming mean ellipticity 0.35, in
0.5 dex intervals. The continuous line represent the OLSB fit for the giants.
}\label{kormendy}
\end{figure}
\clearpage

The ``Kormendy relation'' \citep{Kormendy85} between the effective surface brightness
\mue and the effective radius \re is shown in Fig.\ref{kormendy}. Filled symbols
represent giant ellipticals, empty symbols dwarfs. The dotted lines indicate
iso-luminosity loci, assuming a mean ellipticity of 0.35, from $10^8$ to $10^{12} L_{H\odot}$.
Giant ellipticals follow
a mean relation given by \mue=$(2.84\pm0.06)\cdot \mathrm{Log}~R_{\rm e} + (15.65\pm0.03)$. This is
obtained from an Orthogonal Least Square Bivariate (OLSB) fitting procedure with 3-$\sigma$
rejection, and is represented in Fig.\ref{kormendy} by the continuous line.
Dispersion is noticeable: 0.62~\marc. The slope ($-0.88\pm0.02$ 
expressed in $\mathrm{Log}~ (L_{H\odot}~kpc^{-2}) / \mathrm{Log}~kpc$) is consistent with 
$-0.83\pm0.08$ as measured by \cite{KormendyDjorgowski} in the {\it r} band.\\
Low-luminosity and dwarf ellipticals do not follow the same relationship as giants:
the latter would imply that galaxies at lower luminosity should be increasingly compact, M32-like
objects, while observations show that the lower luminosity of dwarfs is mainly
due to their lower surface brightness. However, from Fig.\ref{kormendy} no clear
dichotomy can be assessed between giants and dwarfs, and it could be argued
that, to a first approximation, galaxies populate uniformly the \mue-\re space, 
avoiding the region below the
mean Kormendy relation. In other words, for each effective radius there is an 
upper-limiting effective surface brightness, and this is lower for larger radii. 
\section{The FP relation}\label{FP_sect}
\begin{figure}
\plotone{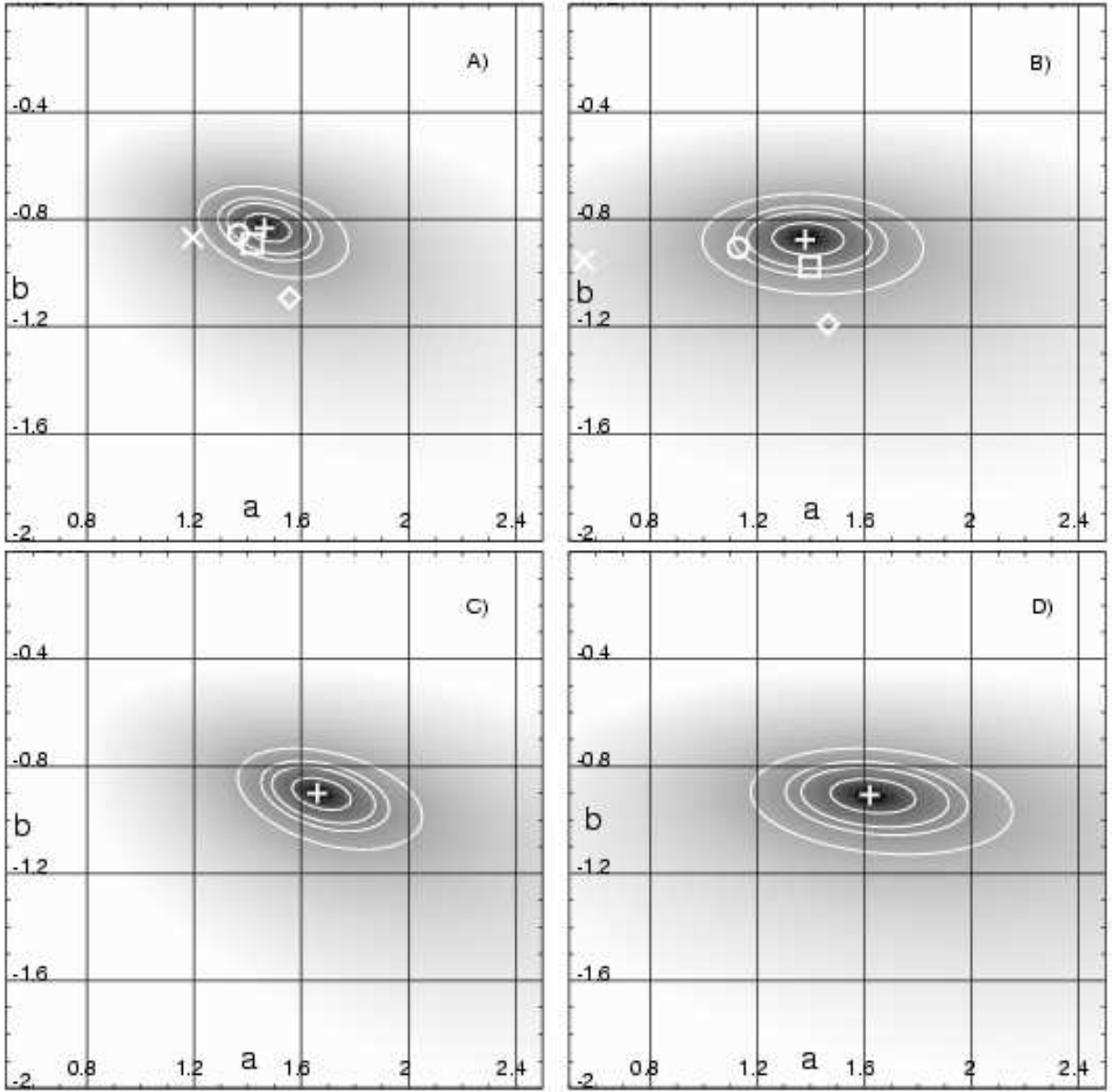}
\caption{The $\Delta\chi^2(a,~b)$ distribution with respect to the minimum 
in logarithmic grey scale for the OEW+correlation fit on the whole sample (A)) and 
on the sample having $\rm \sigma_0~>~100~km~sec^{-1}$ (B)). Minimum is marked by a +.
Confidence levels of 68.3, 
95.4, 99.0 and 99.99\% are shown as superposed contour lines. Superposed symbols
in panel A) and B) mark best MIST fits for BIS with error correlation (X), Y3 
with error correlation (diamond), BIS without correlation (O) and Y3 without 
correlation ($\square$). Panel C) and D) are the same as A) and B) respectively
but using velocity dispersions obtained by the dynamical model (see Sec.
\ref{dyn_model}) in the hypothesis of constant $M/L=1$.
}\label{FPfit}
\end{figure}

In this section we present a study of the Fundamental Plane relationship
between $R_e$, $I_e$ (defined by $\mathrm{Log}~I_e/L_{H\odot}~kpc^{-2}=16.0-0.4\cdot~\mu_e$ )
and the central velocity dispersion $\sigma_0$. This is described by:
\begin{equation}
\mathrm{Log}~R_e=a\cdot \mathrm{Log}~\sigma_0 + b\cdot \mathrm{Log}~I_e + c
\end{equation}
Many methods are available for inferring the values of $a,~b,~c$. We have considered
the {\it Orthogonal Error Weighted} (OEW) and the {\it Measurements errors and 
Intrinsic Scatter} \citep[MIST, see][]{MIST} fitting procedures.
The OEW method (developed by M.~Scodeggio, S.~Zibetti and P.~Franzetti) is based on the 
minimisation of the error--weighted distance of the measured points from the plane.
Details on the mathematical treatment and a full description of the method are
given in Appendix \ref{OEWfit}. The MIST method requires an ``a priori'' determination
of the covariance matrix for error and intrinsic scatter. The first one has been
estimated as the mean error covariance matrix for the sample. Lacking any reliable model
for the intrinsic scatter, the correspondent covariance matrix has been defined null.
We consider the results for a number of different fitting procedures and parameters,
and separately for the whole dynamical sample (135 galaxies) and for the subsample of 120 
galaxies having $(\rm \sigma_0>100~km~sec^{-1})$\footnote{Values of $\rm \sigma_0<100~km~
sec^{-1}$ are usually affected by noticeable uncertainties.}:
they are presented in Table \ref{fittab} 
and in the panels A) and B) of Fig.\ref{FPfit}. The columns in Table \ref{fittab} are as 
follows:\\
(1) fitting method: MIST or OEW;\\
(2) type of MIST fit: bisector (BIS) or using Log\re as dependent variable (Y3);\\
(3) error correlation for Log~$I_e$-Log~\re: if {\it yes} 
$\sigma_{\mathrm{Log}~R_e~\mathrm{Log}~I_e}^2\sim 0.015\sim 1.2~\sigma_{\mathrm{Log}~R_e}\cdot~\sigma_{\mathrm{Log}~I_e}$ 
is assumed; if {\it no} errors on Log$I_e$ and Log\re are considered 
uncorrelated\footnote{this is the case if the photometric calibration error is the
main source of uncertainty on Log$I_e$.};\\
(4) lower limit on $\sigma_0$ ($\mathrm{km~sec}^{-1}$);\\
(5)-(6) $a$ and associated 1-$\sigma$ (68.3\% confidence level) error;\\
(7)-(8) $b$ and associated 1-$\sigma$ (68.3\% confidence level) error.\\

\clearpage
\begin{table}[ht]
\caption{Fit parameters.}\label{fittab}
\[
\begin{array}{cccccccc}
\hline
MIST/OEW &  type &  corr & \sigma_0 & a & err & b  & err \\  
(1)  & (2)  & (3) & (4)  &  (5)   & (6)  & (7) & (8)\\
\hline
MIST	&BIS	&yes	&0	&1.20 	&0.11   &-0.87  &0.04   \\
MIST	&Y3	&yes	&0	&1.56 	&0.13   &-1.09  &0.08   \\
MIST	&BIS	&no	&0	&1.37 	&0.09   &-0.86  &0.04   \\
MIST	&Y3	&no	&0	&1.42 	&0.10   &-0.89  &0.06   \\
MIST	&BIS	&yes	&100	&0.56 	&0.12   &-0.95  &0.05   \\
MIST	&Y3	&yes	&100	&1.47 	&0.15   &-1.19  &0.10   \\
MIST	&BIS	&no	&100	&1.13 	&0.12   &-0.91  &0.04   \\
MIST	&Y3	&no	&100	&1.40 	&0.12   &-0.98  &0.06   \\
OEW     &-      &yes    &0      &1.46   &0.1    &-0.83  &0.07   \\
OEW     &-      &no     &0      &1.47   &0.1    &-0.84  &0.07   \\
OEW     &-      &yes    &100    &1.38   &0.1    &-0.88  &0.07   \\
OEW     &-      &no     &100    &1.41   &0.1    &-0.87  &0.07   \\
\end{array}
\]
\end{table}
\clearpage

Inferred parameters $a,~b$ show very large inconsistencies in the case of MIST fit,
depending on the sample selection, on the error correlations and on the type of fit,
bisector or with an individual dependent variable.
On the other hand, OEW fits are more stable and consistent within the error limits.
This is illustrated in the panels A) and B) of fig.\ref{FPfit}.

\clearpage
\clearpage

\begin{figure}
\plotone{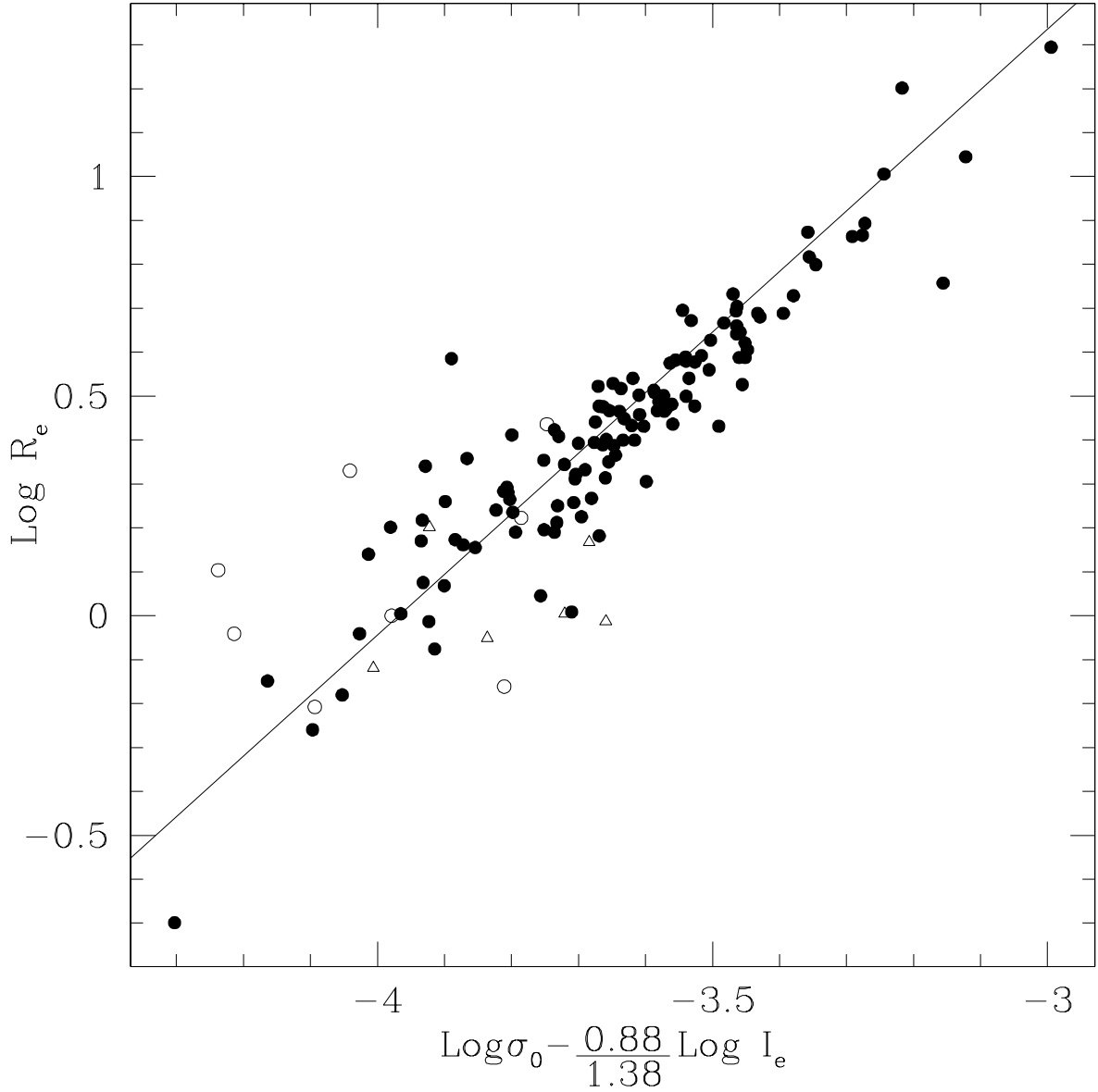}
\caption{A FP section showing Log~\re distribution along the fitted plane. Triangles
identify dwarf galaxies, circles giants. Open symbols are for galaxies having 
$\rm \sigma_0<100~km~sec^{-1}$. Typical error bars are 0.09 dex along Log~\re axis and
0.10 dex along the x-axis.
}\label{FPsection}
\end{figure}

The distribution
of the $\Delta\chi^2(a,~b)$ with respect to the minimum (marked by a +) is represented 
in logarithmic grey scale for the OEW+correlation fit on the whole sample (A)) and 
on the sample having $\rm \sigma_0>100~km~sec^{-1}$ (B)). Confidence levels of 68.3, 
95.4, 99.0 and 99.99\% are shown as superposed contour lines. Superposed symbols
mark best MIST fits for BIS with error correlation (X), Y3 with error correlation 
(diamond), BIS without correlation (O) and Y3 without correlation ($\square$).
MIST fit results are substantially inconsistent with the OEW results as well, being
inside the 95.4\% confidence level at best, and outside 99.99\% confidence level
when error correlation is considered. These inconsistencies are probably due to
the high sensitivity of the MIST algorithms to the error and intrinsic scatter model,
which is ill-determined.\\
As our best estimate of the fundamental plane in the H pass-band we adopt the OEW fit with 
error correlation as applied to the $\rm \sigma_0>100~km~sec^{-1}$ limited sample:
\begin{equation}
\mathrm{Log}~R_e=(1.38\pm0.1)\cdot \mathrm{Log}~\sigma_0 -(0.88\pm0.07)\cdot \mathrm{Log}~I_e + 5.47
\end{equation}
In fig.\ref{FPsection} we show the Log~\re distribution along the fitted plane.
The average dispersion is 0.14, which is somewhat higher than the average 1-$\sigma$
error associated with Log~\re (0.09). We conclude that intrinsic scatter is present in
the FP relation.

\clearpage
\clearpage

\section{Origins of the FP tilt}\label{tilt_sect}
\begin{figure}
\plotone{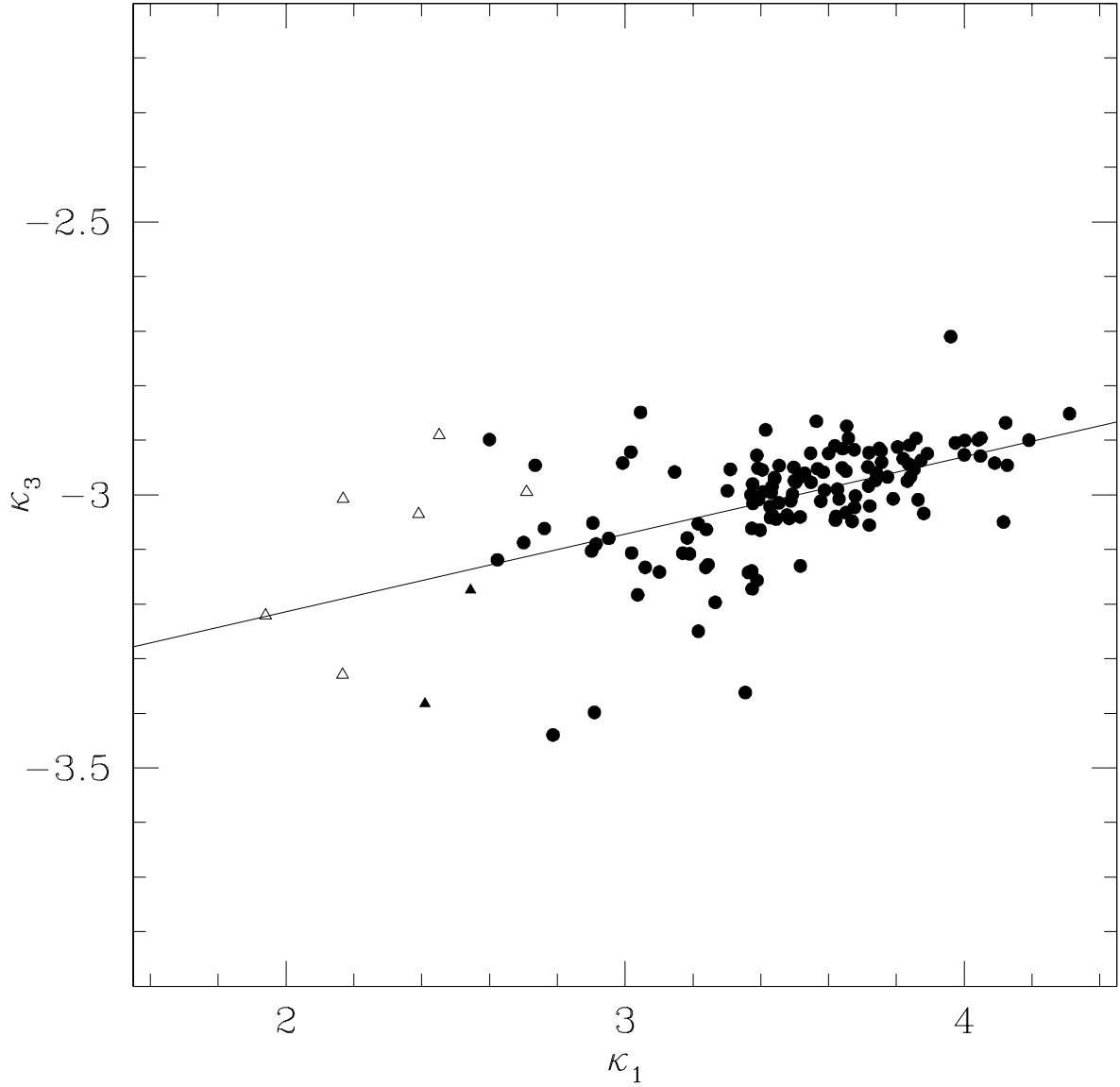}
\caption{ The $\kappa_3-\kappa_1$ projection. Triangles
identify dwarf galaxies, circles giants. Open symbols are for galaxies having 
$\sigma_0<100~km~sec^{-1}$. The continuous line is the best least square fit:
$\kappa_3=0.14\cdot\kappa_1-3.50$.
}\label{k1k3}
\end{figure}

The zeroth order interpretation of the origins of the FP relationship is derived
applying virial equilibrium and the hypothesis that elliptical galaxies form a
homological family with respect to their structural and dynamical parameters. That implies:
\begin{equation}
M\cdot\frac{1}{R_g}\propto\left<v^2\right>,~R_g\equiv\frac{GM^2}{U}
\end{equation}
\begin{eqnarray}
\lefteqn{M\propto\frac{M}{L}\cdot~I_e\cdot~R_e^2}\\
\lefteqn{R_g\propto R_e}\\
\lefteqn{\left<v^2\right>\propto\sigma_0^2}\label{dyn_homology}
\end{eqnarray}
where $M$ is the total mass, $R_g$ is the gravitational radius,
$\left<v^2\right>$ is the mean squared velocity of the ``particles'' (i.e. the stars),
$G$ is the gravitational constant and $U$ is the total gravitational potential energy.
So the expected equation for the Fundamental Plane is:
\begin{equation}
\mathrm{Log}~R_e=2\cdot \mathrm{Log}~\sigma_0 - \mathrm{Log}~I_e - \mathrm{Log}~\frac{M}{L} + k\label{virialFP}
\end{equation}
where the $k$ constant includes all the ``shape coefficients'' used for translating 
the observed
quantities into the virial ones, as well as the unit conversion factors.
According to the homological hypothesis, the difference between the measured 
coefficients and the expected ones (the so called ``tilt'' of the FP) is entirely
produced by a systematic change in  $M/L$ along the galaxy luminosity sequence.
This is shown for our sample in fig. \ref{k1k3}, using the ``$\kappa$ space'' 
formalism \citep{k3} where:\\
\begin{eqnarray}
\kappa_1=(\mathrm{Log}~\sigma_0^2 + \mathrm{Log}~R_e)/\sqrt{2}\propto~\mathrm{Log}~M\\
\kappa_3=(\mathrm{Log}~\sigma_0^2 - \mathrm{Log}~\Sigma_e - \mathrm{Log}~R_e)/\sqrt{3}\propto~\mathrm{Log}~\frac{M}{L}
\end{eqnarray}

\clearpage
\clearpage

A systematic increase in $M/L$ with mass is found, in agreement with the
claims of \cite{k3} in B band and \cite{pahre98b} in K band.\\
However, the analysis of surface brightness profiles of elliptical galaxies performed
in different  pass-bands by many authors during the last years \citep[for the H band see][]{dEvirgo,c31marco} 
demonstrated a systematic change in the profile shapes 
with luminosity and/or radii, requiring some deviation from pure homology. 
Systematic variations of the $k$ term in equation (\ref{virialFP}) have to be taken
into account and $\kappa_3$ no longer represents a trend of $M/L$, but 
results from a combination of $M/L$ and ``shape coefficient'' effects. In order to 
disentangle the two contributions we developed the model described in the next section.
\subsection{Dynamical model}\label{dyn_model}
We model elliptical galaxies as spherically symmetric, isotropic, pressure-supported 
dynamical systems.
Deviations from spherical symmetry and the contribution of rotation to the kinetic energy are
neglected \cite[see][]{ciotti97}.
$M/L$ is left as a free parameter, but is assumed to be constant within each galaxy.
The radial light density profile is derived by de-projection of the measured azimuthally
averaged surface brightness profile. This is translated into the mass density
profile assuming an initial guess value for $M/L=1.0$ (solar units). Solving the hydrostatical
equilibrium equation the velocity dispersion profile is obtained. The contribution
to the Doppler line broadening of each mass element is integrated 
along the line of sight within a characteristic slit aperture to obtain the model
central velocity dispersion $\sigma_{0m}=\sigma_{0m}(M/L)$. $M/L$ is then 
adjusted to match the measured $\sigma_0$. It is worth noticing here, that the computed values
of $\sigma_0$ are influenced both by the intrinsic non-homology of the galaxies (i.e. different shapes of
mass distribution profiles) and by the aperture effects, due to the different fraction of
the galaxy light that contributes to the line broadening inside the slit aperture.\\
Details on the mathematical treatment are given in appendix \ref{app_model}.

\subsection{$M/L$ ratios}
Unlike in the case of the $\kappa_3$ coefficient, the values we derive here
for $M/L$ are not influenced by the varying shape of the surface brightness 
profiles and by the aperture effects.

\clearpage
\begin{figure}
\plotone{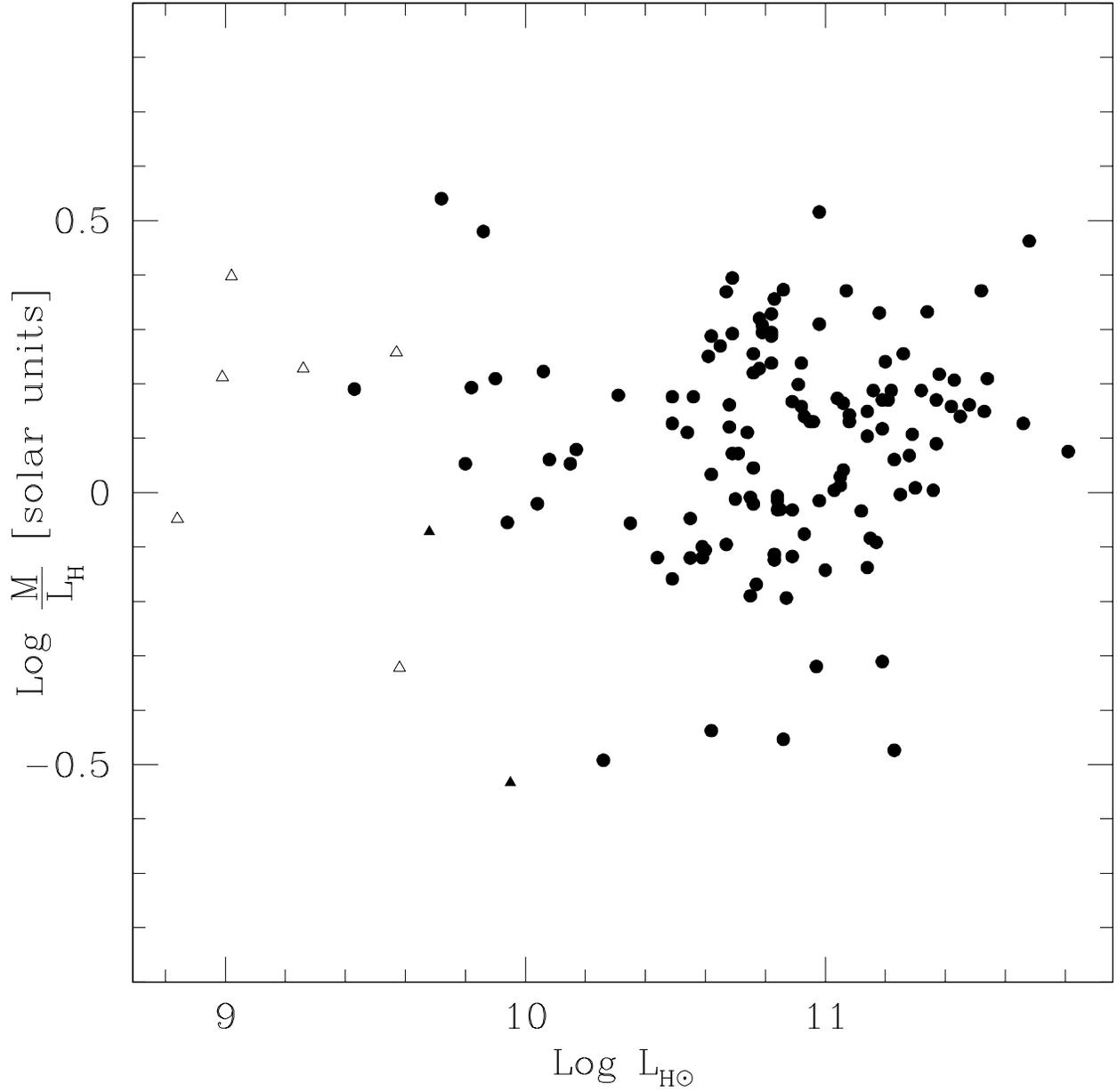}
\caption{ The $M/L$ ratios as derived from the model as a function of the total
H band luminosity. Symbols as in fig.\ref{k1k3}.}\label{m2l}
\end{figure}
\clearpage

In fig.\ref{m2l} Log$M/L$ is shown as a function of the total
H band luminosity. No significant evidence of systematic variation of $M/L$ with $L$ 
can be seen. The 
average Log$M/L$ is 0.09 (corresponding to $M/L\sim1.2$), with a large dispersion of
$0.2~dex$.
Systematic variations of $M/L$ are therefore unlikely to be solely responsible for the tilt
of the FP. Further evidence comes from replacing the measured values of $\sigma_0$ with those
obtained from the model assuming a constant $M/L$ for all galaxies\footnote{The exact
value of $M/L$ is irrelevant in order to obtain the $a$ and $b$ coefficients. 
$M/L=1.0$ has been assumed for simplicity.}: 
the resulting distribution of the 
$\Delta\chi^2(a,~b)$ is shown in the panels C) and D) of Fig.\ref{FPfit} 
for the whole dynamical sample and
for the subsample with $\sigma_0>100~\mathrm{km~sec}^{-1}$ respectively. 
In both cases the distribution is
inconsistent with the values $a=2,~b=-1$ expected on the basis of the virial theorem 
and homology. This demonstrates that the observed deviations of the profiles from
self-similarity are expected to cause a FP tilt similar to that observed, in the
absence of any variation in $M/L$.
\subsection{Homology breaking and aperture effects}\label{apertures}
\begin{figure}
\plotone{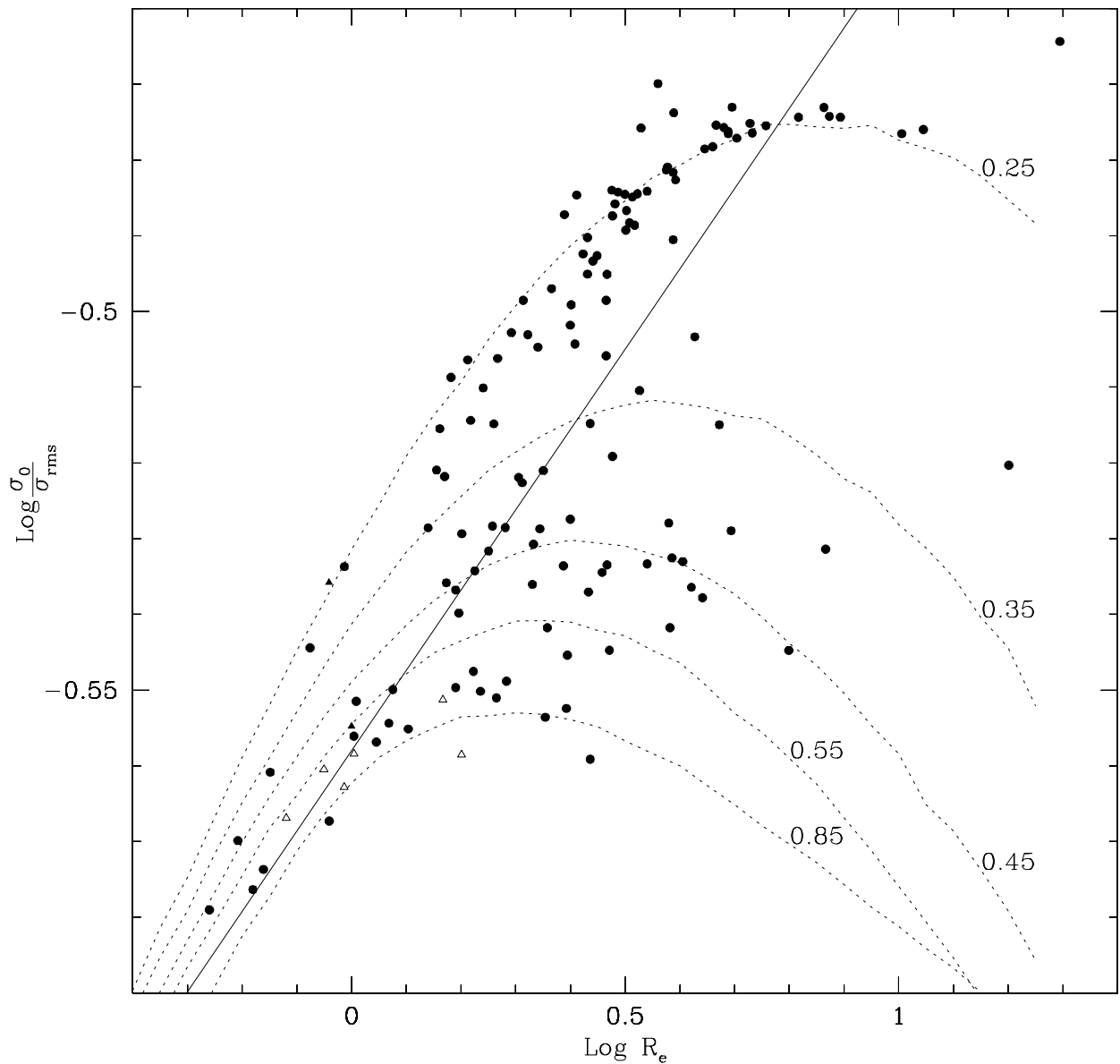}
\caption{The $\sigma_0/\sigma_{rms}$ ratio as a function of \re. Symbols as in 
fig.\ref{k1k3}. The continuous line represents the Orthogonal Least Squares 
Bivariate fit. The dashed lines represent the theoretically expected relation for
families of homologous galaxies modelled by S\'ersic laws: each curve is labelled
with the corresponding $\nu $ index.} \label{sigma_ratio}
\end{figure}

The dynamical model provides the phase-space density distribution for each galaxy. Potential
and kinetic energy can be calculated and the gravitational radius $R_g$ (see equation \ref{rg})
and the rms velocity $\left<v^2\right>^{1/2}\equiv\sigma_{rms}$ can be determined as well.\\
In order to investigate
the consequences of the breaking of the homology and of the aperture effects
on the coefficients of the FP relation,
we study the dependence of the ratios $\sigma_0/\sigma_{rms}$ and $R_e/R_g$
on \re. $R_e/R_g$ does not show any significant correlation with \re. 
The relation between $\sigma_0/\sigma_{rms}$ and \re ~is shown 
in fig.\ref{sigma_ratio}.

\clearpage
\clearpage

The clear correlation
observed between these two parameters demonstrates that the commonly assumed linear proportionality 
between $\sigma_0$ and $\sigma_{rms}$ (eqtn.\ref{dyn_homology}) does not hold. 
In principle, this can be a consequence both of an
aperture effect, depending on the slit aperture relative to \re, and of the different velocity
dispersion profiles.

The Orthogonal Least Squares Bivariate fit for the points gives 
\begin{equation}
\mathrm{Log}\frac{\sigma_0}{\sigma_{rms}}=0.106\cdot\mathrm{Log}R_e-0.558
\end{equation}
and is shown as the solid line. The dashed lines superposed to fig.\ref{sigma_ratio} show the 
theoretically expected aperture effect for
homologous galaxies following the S\'ersic's law\footnote{$I(r)=I_0~Exp\left(-(r/a)^\nu\right)$, 
where $I_0$ is the central surface brightness, $\nu$ is the index ($\nu=0.25$ corresponding to a 
deVaucouleurs profile, $\nu=1$ to an exponential), and $a$ is the scale factor.}
\citep{sersic1968} with index $\nu=0.25,~0.35,~0.45,~0.55,~0.85$. 
Each curve reflects the dependence of the velocity dispersion on the radius and presents a 
linearly increasing part, whose slope is largely independent from $\nu$,
followed by a maximum and a decreasing part, whose location and slope strongly depend on $\nu$.
It is evident that the slope of the mean relation is largely determined by   
the aperture effect in its linear part, while the non-homology of the light profiles (given by the
different $\nu$ index) mostly contributes to the scatter.
Inserting the average relation in the virial FP equation
(\ref{virialFP}), the FP coefficients $a=1.65$ and $b=0.83$ are obtained. Such values
are in very good agreement (within 68.3\% confidence level) with those obtained fitting
the FP on the data sample in which $\sigma_0$ was replaced by the central velocity dispersion
modelled assuming constant $M/L$ (see fig.\ref{FPfit} panels C), D)). The agreement with
the ``real'' FP coefficients is worse, though within the 99.99\% confidence level 
(see Fig.\ref{FPfit} panels A), B)).

\section{Discussion}\label{discussion}
The analysis performed in Sec.\ref{FP_sect} confirms the existence of the
FP relation in the NIR H-band claimed by \cite{FPmarco}. 
The inconsistencies between the plane coefficients obtained
applying different fitting techniques stresses the fact that comparison between
different FP determinations has to be done using homogeneous fitting methods
and, in any case, using large samples of galaxies. Due to its stability, irrespective
of different hypotheses on the covariance matrix, the OEW fit appears to be
the most reliable one.\\
Our determination of FP in H pass-band 
($a=1.38\pm0.1,~b=-0.88\pm0.07,~c=5.47$, see fig.\ref{FPfit}) is consistent
within 1-$\sigma$ with that obtained by \cite{FPmarco} using a subsample of 
73 of our galaxies in the Coma cluster and with that obtained by 
\cite{pahre98b} in K band with a slightly different fitting method.
In our analysis we did not apply any correction for the sample incompleteness bias.
The method used to calculate such corrections \citep[see][ and references therein]{FPmarco}
relies on Monte Carlo simulations to generate a synthetic sample of galaxies reproducing
the observed relations between the galaxy luminosity and the observable quantities involved
in the FP. However, such relations are very poorly determined when extended to the faint 
objects and this affects severely the reliability of the bias corrections. Moreover,
although the present sample represents a significant extension of the one considered by
\cite{FPmarco} toward low luminosities, we find that the values of the FP coefficients
determined in the present work are closer to the uncorrected ($a=1.51\pm0.09,~b=-0.80\pm0.03$) 
than to the corrected ($a=1.66\pm0.10,~b=-0.85\pm0.03$) ones in \cite{FPmarco}, thus 
indicating that the assumptions underlying the bias correction may be incorrect.\\
Applying a dynamical model to the measured surface brightness profiles
the resulting FP tilt and its origins have been studied.
This model represents a higher order approximation than that given by the homological 
hypothesis, as the ``shape coefficients''
translating the measured parameters into the virial ones are allowed to vary according to the
observed surface brightness profile.
Only two assumptions on the structure and dynamics are made:
\begin{itemize}
\item{$M/L$ is constrained to be homogeneous within each galaxy;}
\item{spherical symmetry and isotropic velocity dispersion are assumed.}
\end{itemize}
Real galaxies are actually much more complex systems; nevertheless it is worth noticing that
discrepancies from this description are not likely to affect our conclusions considerably.
The use of NIR pass-band photometry minimises the variation of $M/L$ due to different 
stellar populations. Moreover \cite{ciotti97}
showed that asymmetry and anisotropy affect only slightly the kinetic
energy in elliptical galaxies modelled according to S\'ersic profiles.\\
Contrary to results of \cite{pahre98b} in K band and by several authors
in optical pass-bands \citep[see e.g.][]{k3}, the $M/L$ ratio estimated on the basis
of our dynamical model in H band does not show any significant increase
with luminosity. 
The H band luminosity appears to be a good first-order estimator for the dynamical mass
of elliptical galaxies, in analogy to the claim by \cite{gavrel}, who found direct
proportionality between the H-band luminosity and dynamical mass for disk systems.\\
In turn, the constancy of $M/L$ implies that the tilt of the FP ($\sim56^o$)
is mainly caused by a systematic variation in the laws that translate the 
observed parameters
into the virial ones. In fact, we found that the ratio $R_e/R_g$ does not present
any systematic variation (spatial homology holds), while the ratio 
$\sigma_0/\sigma_{rms}$ increases with
increasing \re. As demonstrated in Sect.\ref{apertures}, this is mainly consequence
of the ``aperture effects'', that strictly related to the way $\sigma_0$ is measured, 
rather than of structural non-homology.
The systematic variation of $\sigma_0/\sigma_{rms}$ can entirely account for the tilt of the 
synthetic FP ($\sim53^o$) in which $\sigma_0$ has been computed from the model 
assuming constant $M/L$. Moreover this leads to the conclusion that the kinetic energy scales
approximately as $\sigma_0^{1.65}$ instead of $\sigma_0^{2}$, in agreement with the 
claim of \cite{busarello97}.
However, some residual systematic variation of $M/L$ 
and $R_e/R_g$ is required to account for the tilt of the measured FP.\\
The analysis performed by \cite{Bertin02} by applying a similar dynamical model
on a sample of 14 galaxies observed in B-band, concluded that, though no strict
homology holds for the elliptical galaxies, a non-negligible contribution
from $M/L$ variation is needed to account for the tilt of the FP in B-band.
However, no aperture effects are considered in this work, and this in turn could lead
to overestimate the contribution due to the $M/L$ variation.\\
The tilt has been proved to increase at shorter wavelengths 
\citep[see][]{FPmarco,pahre98b}. In principle this could be due either to systematic
variations of the global $M/L$ ratio at the shorter optical wavelengths 
or to the different way in which
the aperture effects and non-homology act with respect to the NIR pass-bands, e.g. due to
colour gradients. However, \cite{c31marco}
showed that the colour-magnitude relation $B-H$ vs. $H$ for the ellipticals is
very flat. Given the constancy of $M/L$ in H-band, this in turn rules out large
variations of global $M/L$ as a main cause of the increased tilt of FP at shorter wavelengths.
The influence of the aperture effects and of the profile shapes on the tilt in the 
optical B and V bands will be analysed in a forthcoming paper.\\
The tightness of the FP relation shows that the causes of the
tilt, whatever they are, should act while preserving a structural and dynamical
continuity along the sequence of ellipticals. The structural continuity between
the giant and the dwarf regime is confirmed also by the Kormendy projection 
of the FP (see Sec.\ref{kormendy_sect}). An important point emerging from
the \re-\mue distribution is the existence of a cut-off in the effective
surface brightness: the upper limit in surface brightness decreases with increasing
effective radius, with the same slope of the Kormendy relation for giants.

\section{Summary and conclusions}\label{conclusions}
We have determined the Fundamental Plane of elliptical galaxies
in nearby rich clusters (mainly Virgo and Coma) in NIR
H pass-band. Our result may be written:
$\mathrm{Log}~R_e=(1.38\pm0.1)\mathrm{Log}~\sigma_0 -(0.88\pm0.07)\mathrm{Log}~I_e + 5.47$\\
The relation is tight, showing a dispersion of $0.14~dex$ in \re, while typical
errors are $0.09~dex$.\\
The origins of the tilt of the fitted plane with respect to the virial predictions
for a homologous family of galaxies have been investigated by means of a simple
dynamical model. Spherical symmetry, hydrodynamical equilibrium and isotropic
velocity dispersion are assumed. Constant $M/L$ within each galaxy has been
assumed as a free parameter and used to calculate density and velocity 
dispersion profiles. $M/L$ was then adjusted in order to match predicted values
of $\sigma_0$ with the measured ones.\\
The obtained values of $M/L$ do not show any dependence on the total
luminosity. Systematic variations of $M/L$ are then ruled out as the main cause
of the tilt of the Fundamental Plane. 
We showed the ratio between the rms velocity and the central velocity dispersion
is systematically varying as a function of \re, mainly due to the different slit aperture
relative to \re ~in the spectroscopic measurement of $\sigma_0$. This variation is responsible
for most of the amount of the tilt.\\
The constancy of $M/L$ makes the H-band total luminosity a reliable and cheap 
first-order estimator for the dynamical mass of elliptical galaxies. This matches
the analogous claim by \cite{gavrel} for disk galaxies.\\
Applying the dynamical model to datasets extended to low-luminosity and dwarf elliptical 
galaxies, that will become available in the future, will provide crucial informations
in order to determine whether the hypotheses of constant $M/L$ and
pure isotropic pressure support can describe (at least in first approximation) 
the whole family of ellipticals.
The structural continuity between the giant- and the dwarf- regime is shown 
by the distribution of galaxies in the 
\mue-\re ~plane. The existence of an upper-limit to the effective surface brightness
has been found which follows the classical Kormendy relation for the giants.\\

\acknowledgements{We thank Gianni Busarello and Francesco La Barbera for 
kindly providing the MIST code and for helpful advice, and Simon D.~M. White
for the useful discussion.\\
}

\appendix
\section{OEW fit}\label{OEWfit}
The Orthogonal Error Weighted fit minimises the orthogonal $\chi^2$, defined by
the sum of the squared distances from the fitted plane, weighted on the errors.
This writes:
\begin{equation}
\chi^2=\Sigma_i\frac{d^2(i)}{\delta_d^2}
\end{equation}
were distances are defined by
\begin{equation}
d(i)=\sqrt{\frac{\left(a~\mathrm{Log}~\sigma_0(i) +b~\mathrm{Log}~I_e(i)+c-\mathrm{Log}~R_e(i)\right)^2}{1+a^2+b^2}}
\end{equation}
and corresponding errors are
\begin{equation}
\delta_d=\sqrt{\frac{a^2\delta_{\sigma_0}^2+b^2\delta_{I_e}^2+\delta_{R_e}^2
-2~b~\delta_{I_e}\delta_{R_e}r}{1+a^2+b^2}}
\end{equation}
In the last equation $\delta_{\sigma_0},~\delta_{I_e},~\delta_{R_e}$ are the errors
on the logarithm of the corresponding quantities, $\mathrm r$ is the error
correlation factor between $\mathrm{Log}I_e$ and ${\mathrm Log}{R_e}$ 
(see Sec. \ref{FPsection}).
The minimum is approached using a Powell algorithm (see \cite{NR}) starting from an
initial guess value in the 3-space of the FP coefficients obtained combining the linear fits
on the three projections of the plane.

\section{The dynamical model}\label{app_model}
The dynamical model adopted in this paper for the elliptical galaxies assumes:
\begin{itemize}
\item[-]spherical symmetry;
\item[-]isotropic velocity dispersion tensor;
\item[-]hydrostatic equilibrium;
\item[-]homogeneous $M/L$ ratio throughout the galaxy.
\end{itemize}
The light density distribution $\nu(r)$ is obtained de-projecting the azimuthally
averaged surface brightness profile via the inversion of the Abel integral 
(\cite{binneytr}, eqtn. 4.58a):
\begin{equation}
\nu(r)=-\frac{1}{\pi}\int\limits_{r}^{\infty}\frac{dI(R)}{dR}\frac{dR}{\sqrt{R^2-r^2}}
\end{equation}
Due to the inconsistencies (negative light density) induced by irregularities and
slope inversions in the empirical profiles, these are replaced in the calculations
by the profiles fitted according to analytical laws: deVaucouleurs ($r^{1/4}$, 
\cite{deV}), exponential, bi-component (exponential+exponential or exponential+deVaucouleurs)
and exponentially truncated deVaucouleurs or exponential (see \cite{paperV} and \cite{dEvirgo}
for details).\\
Assuming an initial guess value for the $M/L$ ratio, the mass density profile is
obtained:
\begin{equation}
\rho(r)=\frac{M}{L} \cdot \nu(r)
\end{equation}
For isotropic dispersion tensor the equation for the hydrostatic equilibrium may be written:
\begin{equation}
\frac{d}{dr}\left[\sigma_r^2(r)\rho(r)\right]=-\frac{G M(r) \rho(r)}{r^2}
\end{equation}
By solving the latter, we obtain the radial velocity dispersion profile:
\begin{equation}
\sigma_r^2(r)=\frac{G}{\nu(r)}\frac{M}{L}\int\limits_{r}^{\infty}\frac{L(r')\nu(r')dr'}{r'^2}
\end{equation}
Typical dispersion profiles show a maximum which is more peaked and nearer the centre
the more cuspy the density profile is, in agreement with the claims by \cite{binney} and
of \cite{ciotti97} for the deVaucouleurs and different S\'ersic profiles.\\
Integrating the light weighted contributions to the Doppler line broadening inside a
standard slit aperture (corresponding to 2 arcsec by 6 arcsec at the distance of Coma)
the expected central velocity dispersion $\sigma_{0m}$ is obtained.
As $\sigma_{0m}$ goes as $\left(M/L\right)^{1/2}$, $M/L$ is consequently adjusted
in order to obtain $\sigma_{0m}=\sigma_0$ for the galaxies in the
dynamical sample. For the other galaxies an average $M/L\sim 1.2$ is assumed.\\
Given the hypotheses above and the density and velocity dispersion profiles, the
phase space distribution function $f(\vec{r},\vec{v})$is completely determined and 
can be used to obtain the total gravitational potential energy
\begin{equation}
U=-4\pi\int\limits_0^\infty r^2dr\frac{G~M(r)\rho(r)}{r}
\end{equation}
and the total kinetic energy
\begin{equation}
K=\int\int\left(\frac{1}{2}\vert\vec{v}\vert^2\right)\cdot~f(\vec{r},\vec{v})~d^3\vec{r}~d^3\vec{v}
\end{equation}
In turn, this yields:
\begin{eqnarray}
\lefteqn{R_g=\frac{GM^2}{U}} \label{rg}\\
\lefteqn{\sigma_{rms}\equiv\left<v^2\right>^{1/2}=\sqrt{2K/M}}
\end{eqnarray}
All calculations are implemented numerically in a C code written by S.Z. relying on
standard algorithms taken from \cite{NR}. High numerical accuracy is obtained: the
$K/|U|$ ratio differs from the expected virial value of $1/2$ by less than $10^{-4}$ at
most.

\end{document}